\documentclass{elsart}
\usepackage{graphicx}
\usepackage{amssymb,amsmath}

\begin{document}
\journal{J. of Magnetism and Magnetic Materials}
\begin{frontmatter}

\title{Magnetic properties of amorphous Co$_x$Nb$_{100-x}$ alloys produced by mechanical alloying}
\author{K. D. Machado\corauthref{cor1}},
\ead{kleber@fisica.ufsc.br}
\corauth[cor1]{Corresponding author.}
\author{R. C. da Silva},
\author{J. C. de Lima},
\author{T. A. Grandi},
\author{A. A. M. Gasperini},
\author{C. E. M. Campos}
\address{Depto de F\'{\i}sica, Universidade Federal de Santa Catarina, Trindade, Cx. P. 476, 
88040-900, Florian\'opolis, Santa Catarina, Brazil}

\begin{abstract}

Three amorphous Co$_x$Nb$_{100-x}$ alloys, Co$_{25}$Nb$_{75}$, Co$_{57}$Nb$_{43}$ and Co$_{80}$Nb$_{20}$, 
were produced by Mechanical Alloying starting from the elemental powders. 
Their magnetic properties were determined using an alternating gradient force magnetometer 
(AGFM), and the remanent magnetizations, saturation fields and coercive fields were obtained 
from the hysteresis loop. The alloys have a relatively high saturation field, which decreases 
as the composition becomes richer in Co. The coercivity and remanent magnetization reach an optimal 
value around 57\% at.Co, making {\em a}-Co$_{57}$Nb$_{43}$ the hardest magnetic material 
among the three alloys. Further addition of Co produces a soft alloy.
\end{abstract}

\end{frontmatter}

\section{Introduction}

Mechanical alloying (MA) technique \cite{MA} is an efficient means for synthesizing 
crystalline compounds, amorphous alloys and unstable and metastable 
phases \cite{JoaoFeTi,Weeber,Froes,Yavari,SeZn,KleNiTi,CarlosCoSe}. MA 
has also been used to produce materials with nanometer sized grains and alloys whose components have 
large differences in their melting temperatures and are thus difficult to produce using techniques 
based on melting. It is a dry milling process in which a metallic or non-metallic powder mixture 
is actively deformed in a controlled atmosphere under a highly energetic ball charge. The few 
thermodynamics restrictions on the alloy composition open up a wide range of possibilities for 
property combinations \cite{Poole}, even for immiscible elements \cite{Abbate}. 
The temperatures reached in MA 
are very low, and thus this low temperature process reduces reaction kinetics, allowing the 
production of poorly crystallized or amorphous materials.

Co--Nb is one of many systems of interest due to their superior magnetic properties \cite{Zeng}. Its 
equilibrium phase diagram shows three stable phases Co$_7$Nb$_6$, $\alpha$-Co$_2$Nb and 
Co$_3$Nb, and at least one high-temperature alloy, $\beta$-Co$_2$Nb \cite{Massalki}. Besides the 
techniques used to produce Co--Nb alloys, which are described in Ref. \cite{Massalki}, 
Co--Nb amorphous films 
have been formed by ion-beam mixing (IM) of multilayer films \cite{Zeng}. The amorphization range is 
determined to vary from 23 to 80 at.\% Co. Zeng {\em et al}. \cite{Zeng} have also shown that induced growth 
of such amorphous films can be performed by the ion beam assisted deposition (IBAD) technique. Here, 
we have used MA to produce amorphous Co$_{x}$Nb$_{100-x}$ alloys 
starting from the crystalline elemental powders. The chosen compositions were 
Co$_{25}$Nb$_{75}$ ({\em a}-Co$_{25}$Nb$_{75}$), Co$_{57}$Nb$_{43}$ ({\em a}-Co$_{57}$Nb$_{43}$) 
and Co$_{80}$Nb$_{20}$ ({\em a}-Co$_{80}$Nb$_{20}$). 
Their magnetic properties were determined using an alternating gradient force magnetometer 
(AGFM), and the remanent magnetizations, saturation fields and coercive fields were obtained 
from the hysteresis loop. The alloys have a relatively high saturation field, which decreases 
as the composition becomes richer in Co. The coercivity and remanent magnetization reach an optimal 
value around 57\% at.Co, making {\em a}-Co$_{25}$Nb$_{75}$ the hardest magnetic material 
among the three alloys. If the amount of Co is even increased, the alloy becomes a soft magnetic 
material.

\section{Experimental Procedures}

Binary mixtures of Co and Nb crystalline elemental metals powders, with nominal composition 
Co$_{25}$Nb$_{75}$, Co$_{57}$Nb$_{43}$ and Co$_{80}$Nb$_{20}$ were sealed together with several steel 
balls, under argon atmosphere, in steel vials. The weight ratio of the ball to powder was 6:1 for 
all compositions. The vials were mounted in a Spex 8000 shaker mill and the samples were milled for 
40 h. In order to keep the vial temperature close to room temperature, a ventilation system was used. 
The compositions of the as-milled alloys were determined by X-ray 
fluorescence method using an EDX-700 Shimadzu equipment, giving a composition of 
25 at.\%Co and 75 at.\%Nb for the first alloy, 
57 at.\%Co and 43 at.\%Nb for the second, and 
79 at.\%Co and 21 at.\%Nb for the third. Impurity traces were not detected. 
The hysteresis loops were obtained at room temperature using a home-made alternating gradient force 
magnetometer by applying the magnetic field, up to $\pm 5.0$ kOe, parallel to the plane of sample.

\section{Results and Discussion}

\begin{figure}
\begin{center}
\includegraphics{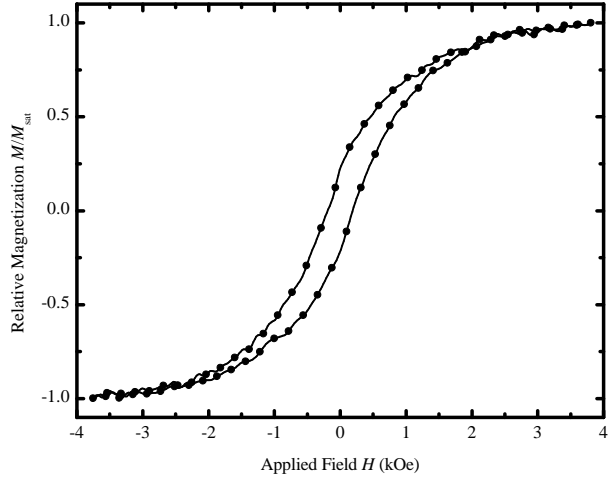}
\end{center}
\caption{\label{fig1} Hysteresis loop of {\em a}-Co$_{25}$Nb$_{75}$.}
\end{figure}

The hysteresis loop of {\em a}-Co$_{25}$Nb$_{75}$ showing the relative magnetization 
$M/M_{\rm sat}$ as a function of the applied field $H$ can be seen in fig. \ref{fig1}. 
From this figure, 
we found the remanent magnetization $M_{\rm rem}$, 
which is about 22\% of the saturation value ($M_{\rm sat}$), and the 
saturation ($H_s$) and coercive ($H_c$) fields are $H_s = 3809$ Oe and $H_c = 183$ Oe, respectively.

\begin{figure}[h]
\begin{center}
\includegraphics{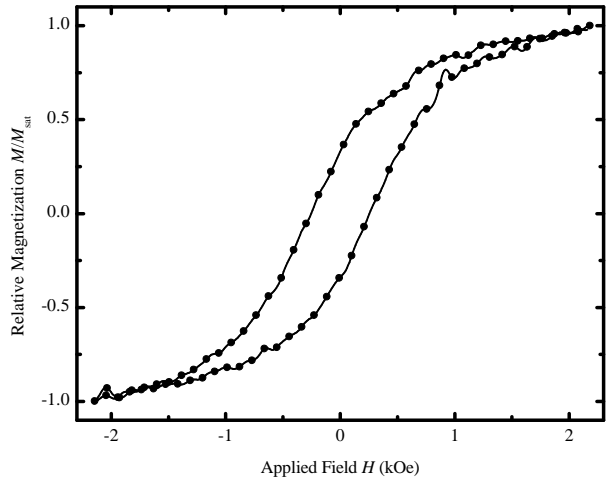}
\end{center}
\caption{\label{fig2} Hysteresis loop of {\em a}-Co$_{57}$Nb$_{43}$.}
\end{figure}

Figure \ref{fig2} shows the hysteresis loop of {\em a}-Co$_{57}$Nb$_{43}$. From this figure it can be seen a 
reduction in the saturation field, which is now $H_s = 2144$ Oe, and an increase both in coercivity 
($H_c = 254$ Oe) and in remanent magnetization, which is 33\% of the saturation value for this alloy. 

\begin{figure}[h]
\begin{center}
\includegraphics{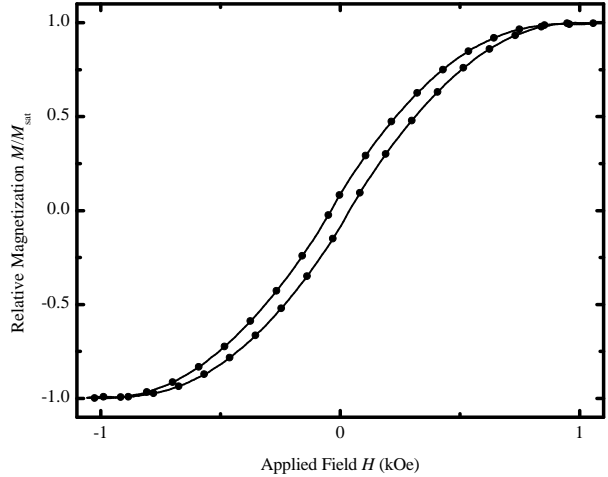}
\end{center}
\caption{\label{fig3} Hysteresis loop of {\em a}-Co$_{80}$Nb$_{20}$.}
\end{figure}

The hysteresis loop of the third alloy, {\em a}-Co$_{80}$Nb$_{20}$, is seen in fig. \ref{fig3}. From this 
figure, we found $H_s=1058$ Oe for the saturation field, which shows again a reduction when compared with 
the values found for {\em a}-Co$_{25}$Nb$_{75}$ and {\em a}-Co$_{57}$Nb$_{43}$, but its coercivity decreases 
to $H_c=35$ Oe and only 8\% of the magnetization is retained from its saturation value. Although the 
addition of Co to the {\em a}-Co$_{25}$Nb$_{75}$ alloy  has produced a harder magnetic material, the 
increase in the magnetic properties seems to reach an optimal value around 57\%at.Co. After that, the 
amorphous Co$_x$Nb$_{100-x}$ alloys become a softer magnetic material, as shown by the data found for 
{\em a}-Co$_{80}$Nb$_{20}$. Table \ref{tab1} shows the data obtained for the three alloys.

\begin{table}
\caption{\label{tab1} Magnetic parameters obtained for amorphous Co$_x$Nb$_{100-x}$ alloys.}
\begin{center}
\begin{tabular}{cccc}\hline
$x$ (\% at.Co) & $H_s$ (Oe) & $H_c$ (Oe) & $M_{\rm rem}/M_{\rm sat}$ (\%)\\
25 & 3809 & 183 & 22 \\
57 & 2144 & 254 & 33 \\
80 & 1058 & 35  & 8  \\\hline
\end{tabular}
\end{center}
\end{table}

\section{Conclusion}

Amorphous Co$_{x}$Nb$_{100-x}$ alloys were prepared by MA and their magnetic properties investigated. 
The main conclusions of this study are:

\begin{enumerate}
\item Formation of {\em a}-Co$_{x}$Nb$_{100-x}$ by MA in the range $25 \le x \le 80$ was 
confirmed experimentally. 

\item Amorphous Co$_{x}$Nb$_{100-x}$ alloys 
are magnetic materials, showing relatively high saturation fields, 
which decreases as the alloy composition becomes richer in Co.

\item Coercivity and remanent magnetization increase when composition goes from 25\% at.Co to 
57\% at.Co, making the {\em a}-Co$_{57}$Nb$_{43}$ the hardest magnetic material among the three alloys. 
However, further increase in the amount of Co atoms did not produce a harder magnetic alloy. $H_c$ and 
$M_{\rm rem}$ decrease and {\em a}-Co$_{80}$Nb$_{20}$ becomes the softest alloy.
 
\end{enumerate}

\ack

We thank the Brazilian agencies CAPES and CNPq for financial support.

%\bibliographystyle{elsart-num}
%\bibliography{conbmag}

\end{document}